\documentclass[aip,reprint]{revtex4-1}

\usepackage[dvips]{graphicx}

\begin{document}

\title{Electrical properties of III-Nitride LEDs: recombination-based injection model and theoretical limits to electrical efficiency and electroluminescent cooling}


\author{Aurelien David}
\email{adavid@soraa.com}
\affiliation{Soraa Inc., 6500 Kaiser Dr. Fremont CA 94555, USA}
\author{Christophe A. Hurni}
\affiliation{Soraa Inc., 6500 Kaiser Dr. Fremont CA 94555, USA}
\author{Nathan G. Young}
\affiliation{Soraa Inc., 6500 Kaiser Dr. Fremont CA 94555, USA}
\author{Michael D. Craven}
\affiliation{Soraa Inc., 6500 Kaiser Dr. Fremont CA 94555, USA}

\date{\today}

\begin{abstract}
The current-voltage characteristic and ideality factor of III-Nitride quantum well light-emitting diodes (LEDs) grown on bulk GaN substrates are investigated. At operating temperature, these electrical properties exhibit a simple behavior. A model in which only active-region recombinations have a contribution to the LED current is found to account for experimental results. The limit of LED electrical efficiency is discussed based on the model and on thermodynamic arguments, and implications for electroluminescent cooling are examined.
\end{abstract}

\pacs{}

\maketitle 
 
The current-voltage ($I$-$V$) characteristic of a light-emitting diode (LED), and the corresponding ideality factor ($n$), are basic electrical properties, yet they remain poorly understood in III-Nitride LEDs. In particular, a simple analytical model to interpret $I$-$V$s is lacking.

From a theoretical standpoint, various models of III-Nitride LEDs overestimate $V$ by hundreds of mV compared to experiments,\cite{Kim07,Kisin11,Piprek13} indicating that the physical transport phenomena are not properly accounted for.\cite{Karpov11}

From an experimental standpoint, there has been focus on understanding the high $n$ (often on the order of 4-8 and as high as 14)\cite{Zhu09,Aufdermaur14} observed in III-Nitride LEDs: these have been justified by inefficient transport across quantum barriers\cite{Zhu09} and by trap-assisted transport.\cite{Aufdermaur14} Recently, we reported on high-efficiency LEDs\cite{Hurni15} and noted that their voltage was extremely low: it remained below the energy of the emitted photons even at high current densities $\sim 100$~A cm$^{-2}$. We noted that this was possible due to the absorption of phonons from the lattice, but that it was somewhat surprising that this effect remain pronounced at high power. Finally, Ref.~\onlinecite{Xue15} recently observed an unexpected increase of wall-plug efficiency at high current with temperature.

This imperfect understanding of the electrical characteristics of III-Nitride LEDs stands in contrast to the simpler case of $pn$ and $pin$ homojunctions, which are well described by conventional drift-diffusion theory. It has been shown in recent years that near-ideal behavior\cite{Sah57} can be observed in $pn$ GaN homojunctions, provided that the material has low defect density; this has been achieved by optimizing MBE growth,\cite{Hurni10} or by employing low-defect bulk GaN substrates.\cite{Hu15}

In this letter, we propose a simple analytical model for an LED's $I$-$V$ and $n$, in the case where active region recombinations dominate over injection effects. We show that this model accurately predicts the behavior of III-Nitride LEDs at operating temperature, and therefore that such LEDs operate at the limit of ideal transport.

As a test vehicle for this discussion, a simple LED test structure was grown by metal-organic chemical vapor deposition on a freestanding bulk c-plane GaN substrate. It consists of an n-doped GaN region, an undoped region containing five InGaN quantum wells (emitting near 435~nm) separated by GaN barriers, and a p-doped GaN region. To simplify transport processes, the structure does not include an electron blocking layer (EBL). We chose this plane orientation and wavelength as representative of standard commercially-available LEDs.

\begin{figure}[!!!tttth]
\centering
\includegraphics[width=8.5cm]{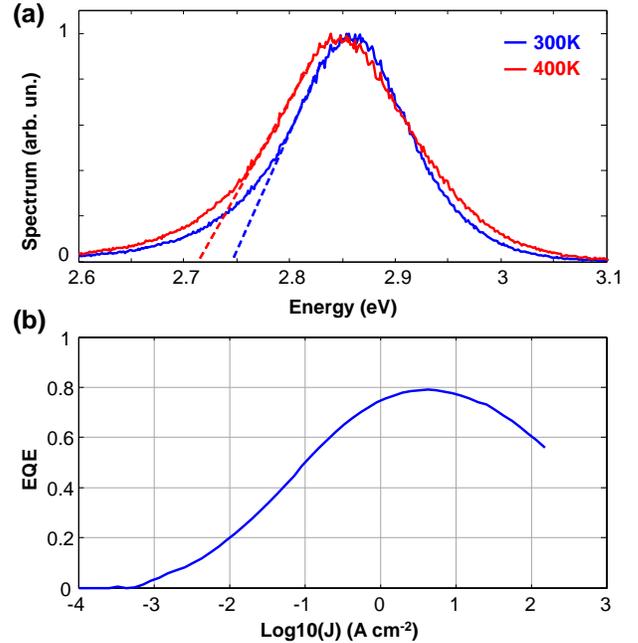} 
\caption{(a) Luminescence spectra at various temperatures, at a current density $J=$1~A cm$^{-2}$. Dashed lines: extrapolation of the optical bandgap $E_o$ from the slope of the main luminescence peak (the low-energy tails correspond to phonon-assisted recombinations and are hence ignored in this procedure). (b) External quantum efficiency versus $J$ at 300~K.}
\label{Fig1}
\end{figure}

Fig.~\ref{Fig1} shows experimental optical properties of the LED: its external quantum efficiency (EQE) and emission spectra. Despite the absence of an EBL, this structure retains a very high efficiency: at 300~K, it peaks at $80\%$. After normalizing for extraction efficiency and package efficiency,\cite{David14a} this corresponds to a peak internal quantum efficiency of about $89\%$.

\begin{figure}[!!!!!!!!!!ttttttth]
\centering
\includegraphics[width=8.5cm]{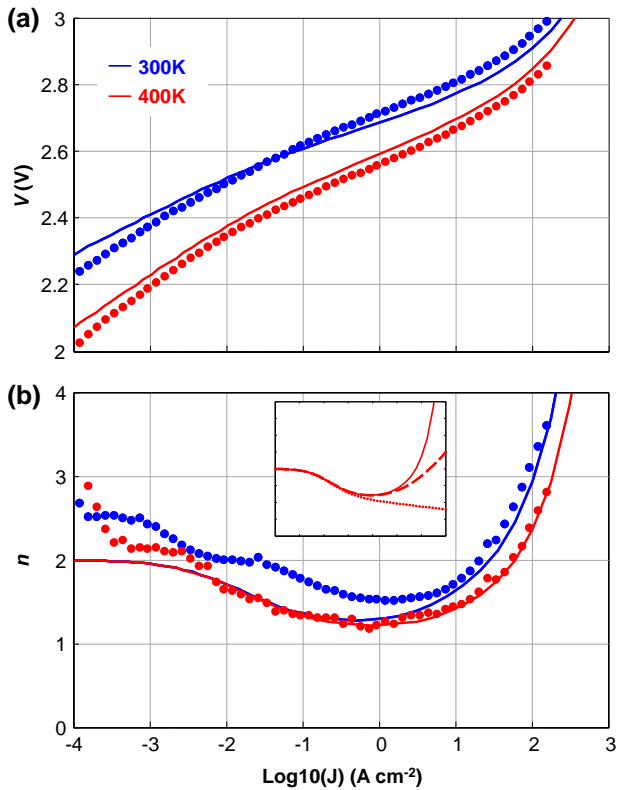}
\caption{(a) Voltage and (b) ideality factor as a function of current density and temperature. Dots: measurements; lines: model. The inset illustrates the ideality factor for different models---dotted: BS (no $R_s$ term); dashed: FDS (no $R_s$ term); full: FDS ($R_s$ included).}
\label{Fig2}
\end{figure}

Fig.~\ref{Fig2} shows electrical properties of the LED: its $I$-$V$ characteristic (measured in 4-point-probe configuration) and $n$. At 400~K, $n$ reaches low values (varying from 2 at low current to about 1 before increasing again). The contrast between these results and previously-reported high $n$ values might be attributed to the high quality of the present material, grown on bulk GaN substrates: the lower level of defects may preclude the defect-assisted transport mechanisms discussed in Ref.~\onlinecite{Aufdermaur14}. Such low $n$ values are somewhat reminiscent of the behavior of an ideal $pn$ junction. This motivates us to develop a model to account for the observed electrical data.

In general, the $I$-$V$ characteristic is controlled by a combination of transport effects and carrier recombinations. Carrier transport involves various physical phenomena---such as standard drift and diffusion, thermionic emission, tunneling, contributions from disorder---whose theoretical description is complex, and remains a challenge even for advanced simulation tools.\cite{Aufdermaur15}

For simplicity, we therefore consider ideal injection, where transport processes have no deleterious impact on the voltage of the LED. In other words, we imagine an LED whose epitaxial design and doping profiles have been optimized such that no voltage drop occurs as carriers are transported across heterostructures and into the active region (except, potentially, for a trivial resistive contribution). Of course, this does not mean that the existence of the microscopic phenomena enabling transport is discarded. Rather, we assume those to be efficient enough to transport carriers into the active region without having a significant impact on the $I$-$V$ characteristic. 

In this limiting case, active region recombination is the only process controlling current through the LED. The $I$-$V$ characteristic can thus be obtained by the following general procedure: relate the Fermi levels to the carrier density (with a model of the quantum density of states); relate the current to the carrier density (with a recombination model); and combine the two relationships. These steps can be achieved by numerical modeling.

Since we seek an analytical model however, we make additional simplifying assumptions. First, we assume that the active region is characterized by unique and uniform quasi-Fermi levels $E_{Fc}$ and $E_{Fv}$ for electrons and holes. This amounts to assuming that the active region is uniformly populated with carriers, which we believe to be the case in these high-quality LEDs. The quasi-Fermi level splitting in the active region is then related to $V$ by: $(E_{Fc}-E_{Fv})=qV$, where $q$ is the elementary charge. Further, we ignore the complexities of the density of states arising from quantum confinement; we simply assume parabolic bulk-like densities of states, with effective densities $N_c$ and $N_v$ in the conduction and valence bands. Introducing the ground state energies $E_c$ and $E_v$ in the conduction and valence bands of the active region, the densities of electrons $N$ and holes $P$ in the active region are, in the Boltzmann limit:

\begin{equation}
\label{eq:model1}
\hspace{-2mm} N = N_c \exp{\left( \frac{E_{Fc}-E_c}{k_BT} \right)},P = N_v \exp{\left(\frac{E_v-E_{Fv}}{k_BT} \right)}
\end{equation}

We now introduce the optical band gap $E_o=E_c-E_v$. Due to quantum confinement and the quantum-confined Stark effect, $E_o$ is in general not equal to the bulk band gap of the active region material. Eq.~\ref{eq:model1} yields:

\begin{equation}
\label{eq:model2}
NP = N^2 = \underbrace{N_cN_v\exp{\left(-\frac{E_o}{k_BT}\right)}}_{N_i^2}\exp{\left(\frac{qV}{k_BT}\right)}
\end{equation}

Eq.~\ref{eq:model2} resembles the semiconductor mass-action law---with $E_o$ replacing the material band gap---and relates the carrier density to the applied voltage.\footnote{This expression ignores the intrinsic carrier density (which is very small in III-Nitrides). It is therefore not valid at extremely low carrier density.}

To relate the carrier density to the current density requires a recombination model. It has been widely documented, especially from carrier lifetime measurements,\cite{Eliseev99,David10a,David10b,Galler12} that recombination rates in III-Nitride LEDs could be well described by the well-known $ABC$ recombination model, in which the current density is $J=qt(AN+BN^2+CN^3)$ with $t$ the active region thickness. By combining  this with Eq.~\ref{eq:model2}, we obtain an analytical model for the current-voltage characteristic:

\begin{eqnarray}
\label{eq:model_final}
\nonumber
\frac{J}{qt} &=& A N_i \exp{\left( \frac{1}{2} \times \frac{qV}{k_BT} \right)} +
B N_i^2 \exp{\left(1 \times \frac{qV}{k_BT} \right)} \\* & + &
C N_i^3 \exp{\left(\frac{3}{2} \times \frac{qV}{k_BT} \right)}
\end{eqnarray}

Each process has a distinct contribution to the $I$-$V$ characteristic and a distinct $n$: 2 for Shockley-Reed-Hall (SRH) recombinations (the same result as in $pn$ junctions), 1 for radiative recombinations, and 2/3 for Auger recombinations. Note that there is no carrier diffusion term in our model. This stands in contrast to conventional $pn$ homojunctions, where the diffusion current is the dominant contribution under operation conditions.

Eq.~\ref{eq:model_final} is intuitive but is limited to moderate carrier density, as it assumes Boltzmann statistics (BS). The model can be easily extended to higher density by considering Fermi-Dirac statistics (FDS). In this case, the $I$-$V$ characteristic no longer has an analytical form but it can still be obtained straightforwardly: at a given carrier density $N=P$, $J$ is given by the $ABC$ model and $V$ by a numerical evaluation of the inverse Fermi-Dirac integral.\cite{Coldren95b}

Evaluating the model requires various numerical parameters. Band structure parameters are taken from Ref.~\onlinecite{Vurgaftman01}. The optical band gap $E_o$ is obtained from the low-energy tail of the spectra of Fig.~\ref{Fig1} (we estimate that this procedure has an accuracy of a few tens of m$eV$, which carries into the accuracy of the voltage prediction). Recombination coefficients are obtained from accurate carrier lifetime measurements.\cite{David15c} We obtain the following values for the active region of the sample under study: $A=1.9 \times 10^5$~s$^{-1}$, $B=3.5 \times 10^{-13}$~s$^{-1}$cm$^{-3}$, $C=7.1 \times 10^{-33}$~s$^{-1}$cm$^{-6}$. Comparison to experiment also requires adding an ohmic term $R_s \cdot I$ to $V$, where $R_s=0.5~\Omega$ is the series resistance.

These parameters yield the fit to data shown in Fig.~\ref{Fig2}. We find that Eq.~\ref{eq:model_final}, which relies on BS, only applies up to $J\sim$1A cm$^{-2}$. Using FDS extends the model's validity to $\sim$100A cm$^{-2}$. Beyond this, the carrier density reaches $10^{19}$~cm$^{-3}$ and high-density effects (bandgap renormalization, field screening) become significant. Interestingly, the effect of FDS is somewhat similar to that of the ohmic contribution (see inset of Fig.~\ref{Fig2}); therefore, inclusion of FDS is necessary for a proper evaluation of $R_s$.

At 400~K, the fit is excellent. This is especially satisfactory since the model contains no free parameters except $R_s$. The model quantitatively predicts the voltage and the ideality factor versus current. Intuitively, the decrease in $n$ from 2 to 1 in the current range $J = 10^{-3}-1$~A cm$^{-2}$ can be traced to the EQE increase in the same current range, from the SRH-dominated regime to the radiative-dominated regime. At high current, observation of the Auger regime ($n$=2/3) is precluded by FDS and the $R_s$ contribution. 

We conclude that in the present sample at high temperature, recombinations indeed control the electrical properties, with transport effects having no measurable contribution. This result is significant in that it indicates that these LEDs essentially operate at the theoretical limit of ideal injection. Remarkably, the band gap of GaN has no bearing on Eq.~\ref{eq:model_final} even though carrier injection proceeds though a GaN matrix. In fact, Eq.~\ref{eq:model2} can also be cast as $qV=E_o+k_B T \ln{(N^2/N_cN_v)}$, which reveals that the optical band gap of the active region determines $V$ at a given $N$. This is in line with the empirical observation that, in well-optimized commercial LEDs, the voltage tends to scale with the photon energy.

At 300~K, $n$ is slightly higher than the model predicts. This points to an additional transport process (possibly imperfect transport across the wells and barriers) causing a departure from ideal injection; the impact of this process disappears at higher temperature. No attempt was made to optimize the epitaxial structure studied here. Further optimization may enable near-ideal injection even at room temperature.

We can now comment on the very efficient injection we reported in Ref.~\onlinecite{Hurni15}. The electrical efficiency (EE) of an LED is defined as $\text{EE}=h \nu / qV$, where $h \nu$ is the average photon energy. This energy is higher than $E_o$ by a few $k_BT$ due to the width of the luminescence spectrum, and can thus be written as $h \nu=E_o+\alpha k_B T$, where $\alpha\sim 5$ characterizes the spectral width. Assuming BS, Eq.~\ref{eq:model2} yields:

\begin{equation}
\label{eq:EE}
\text{EE}=\frac{E_o+\alpha k_B T}{E_o + k_B T \ln{\left(N^2/N_cN_v\right)}}
\end{equation}

Therefore, EE is above unity at low carrier density. Taking FDS into account, we find that EE crosses unity at $J\sim100$~A cm$^{-2}$ (this value of $J$ is roughly at the limit of validity of the model). Therefore the results of Ref.~\onlinecite{Hurni15} are in line with our model: an EE above unity is indeed expected even at high current.

The low voltage predicted and observed in this work stands in contrast to various modeling predictions,\cite{Kim07,Kisin11,Karpov11,Piprek13} in which the forward voltage is much higher than $E_o$, and is actually higher than \textit{the band gap of GaN}---a discrepancy of hundreds of mV. We propose that the present results could be used as a benchmark against which future simulation efforts can be tested. On this topic, recent publications suggest that the high voltage predicted by drift-diffusion models may in part be caused by an improper description of transport across quantum barriers, which can be mitigated by including effective quantum potentials\cite{Bulashevich12} or the effect of alloy disorder.\cite{Yang14} 

Finally, we study the implications of our model regarding electroluminescent cooling. It has long been known that, due to the above-unity value of EE at low enough bias, it is theoretically possible to operate an LED with a wall-plug efficiency (WPE), defined as $\text{WPE}=\text{EQE} \times \text{EE}$, above unity. This may happen if the other factors intervening in power conversion, namely internal quantum efficiency (IQE) and extraction efficiency, are very high. In this regime, the extra energy is provided in the form of heat drawn from the crystal lattice: the LED then acts as a heat pump.\cite{Lehovec53,Tauc57} Microscopically, phonon scattering ensures that the carriers obey a thermal distribution; even at low $V$, high-energy carriers present at the distribution's tails can recombine to create photons at an energy above $E_o$. Experimental proof of this cooling regime, remains elusive (in Ref.~\onlinecite{Santhanam12}, electroluminescent cooling was demonstrated, however in a low-voltage regime which is different from that discussed here). On this topic, we note that the results of Ref.~\onlinecite{Xue15} were discussed in terms of heat absorption; however, the devices studied there had an extremely high room-temperature voltage which collapsed at higher temperature: we believe these poor electrical properties in fact dominate the increase in WPE with temperature of Ref.~\onlinecite{Xue15}.

\begin{figure}[!hhhhhhht]
\centering
\includegraphics[width=8cm]{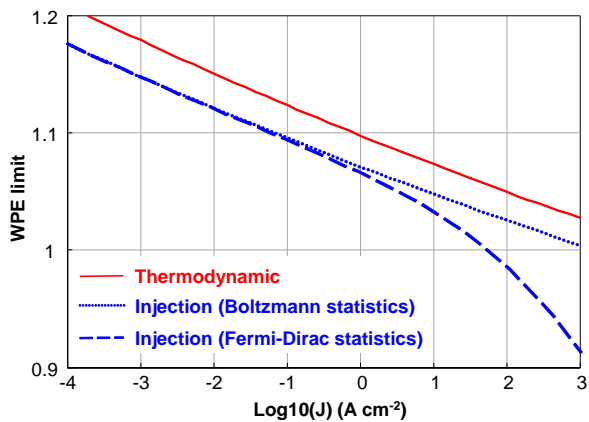}
\caption{WPE limits versus current density, evaluated from various models at $T=300$~K. General thermodynamic arguments yield a limit very close to the injection model with BS.}
\label{Fig3}
\end{figure}

Eq.~\ref{eq:EE} imposes an upper limit for the cooling effect. The highest theoretical WPE would occur in an LED having perfect material without non-radiative recombinations (hence $A=C=0$ and $\text{IQE}=1$), and unity extraction efficiency. In this case, $\text{WPE} = \text{EE}$; in other words, the only limitation to power conversion is the amount of heat that can be drawn by light emission given the statistical carrier distribution. Numerical evaluations of this limit are shown on Fig.~\ref{Fig3}. As already noted, EE is above unity up to high current densities, allowing for cooling.

The limit just derived is based on our detailed injection model; however, another well-known limit is given by general thermodynamics arguments: cooling must comply with the second law of thermodynamics, entropy conservation. Indeed, heat absorbed by the active region increases its entropy whereas emitted photons carry away entropy. The net balance of entropy must remain positive, which limits how many photons can be emitted. This imposes a constraint on the maximum cooling efficiency, discussed in Refs.~\onlinecite{Landau46,Weinstein60,Dousmanis64,Landsberg68,Landsberg80,Wurfel82,Berdahl85}. In short, the LED constitutes a Carnot heat engine, where radiation is the hot reservoir characterized by a temperature $T_B$. Entropy conservation is then expressed by the Carnot limit:

\begin{equation}
\label{eq:entropy}
\text{WPE} \leq \frac{1}{1-T/T_B}
\end{equation}

Eq.~\ref{eq:entropy} imposes an upper limit on WPE. Equality only holds for a perfect material ($\text{IQE}=1$): in the presence of non-radiative recombinations, additional entropy is generated (due to the irreversibility of non-radiative processes). Often termed brightness temperature, $T_B$ is commonly evaluated by imagining that the LED is in equilibrium with a blackbody radiator at the same temperature, emitting at the same intensity in the same spectral range. $T_B$ can then be obtained by numerical integration over the LED's spectrum.\cite{Weinstein60,Landsberg80} For simplicity, we approximate the spectrum as constant in a range $\Delta \nu = \alpha k_B T$ around the central frequency $\nu$ and zero elsewhere. Noting that the radiance of an LED with unity extraction and IQE is $J V/\pi$, this leads to an analytical expression for $T_B$:\cite{Weinstein60}

\begin{equation}
\label{eq:TB}
T_B=\frac{h \nu}{k_B \ln{(1+\rho^{-1})}}, ~\rho=\frac{c^2}{2 h \nu^3} \frac{J V}{\pi \Delta \nu}
\end{equation}

This expression assumes BS. The resulting thermodynamic limit is shown on Fig.~\ref{Fig3}: it leads to values very similar to those of the injection model with BS. This agreement is not coincidental. Indeed, as shown in Refs.~\onlinecite{Berdahl85,Santhanam13}, the brightness temperature can also be directly related to the quasi-Fermi level splitting by: $T_B = T / (1-q V/h \nu) = T / (1-EE)$. Inserting this expression into Eq.~\ref{eq:entropy}, the entropy limit reduces to the definition of electrical efficiency we introduced above. In other words, both limits express the same underlying physics---in essence, this is because radiation from a semiconductor can be derived from thermodynamic considerations. The slight difference shown on Fig.~\ref{Fig3} stems from our numerical approximations. In practice, use  of the injection model enables an evaluation in terms of recombination coefficients (instead of the commonly-used $T_B$), and a straightforward inclusion of the effect of FDS.

This limit has implications for III-Nitride LEDs cooling. At a current density of 1~A cm$^{-2}$ (close to the actual peak IQE in good material), the upper bound of EE is $\sim 1.08$. This illustrates the practical difficulty of achieving cooling in a GaN LED: even if ideal electrical efficiency is achieved, IQE and extraction must remain within a few percent of unity to reach the cooling regime. However, cooling would then occur at a fairly high power density. For instance, if WPE exceeded unity by just 1\%, the cooling power would be on the order of 1~mW cm$^{-2}$ (compared to less than 1~nW cm$^{-2}$ in Ref.~\onlinecite{Santhanam12}).

In summary, we have demonstrated high-performance III-Nitride LEDs grown on bulk GaN having simple electrical characteristics. We have introduced a simple model for the current-voltage characteristic of an LED, in the ideal case where only active region recombinations contribute to injection, and shown that it predicts experimental results at operating temperature. This indicates that these LEDs operate near the theoretical limit of electrical efficiency. Further, we have shown that our model imposes a limit to electroluminescent cooling, which is in fact an alternative form of the well-known entropy-derived limit.

We would like to acknowledge Michael J. Grundmann for early contributions to the recombination-based model, and T. Patrick Xiao for pointing out to us the relationship connecting the two cooling limits.

\bibliography{Biblio_These}

\end{document}